\documentstyle[aps,preprint,epsfig]{revtex}
\tighten
\firstfigfalse

\def\p{{\bbox{p}}}
\def\k{{\bbox{k}}}
\def\q{{\bbox{q}}}
\def\npnk
{\langle \Delta n^\alpha_{\bbox{p}}\, \Delta n^\beta_{\bbox{k}} \rangle}
\begin{document}

\title{Thermal fluctuations in the interacting pion gas}
\author{M. Stephanov}
\address{Department of Physics, University of Illinois, Chicago,
Illinois 60607-7059 and\\
RIKEN-BNL Research Center, Brookhaven National Laboratory,
Upton, New York 11973}
\date{October 2001}

\maketitle

\begin{abstract}
 We derive the two-particle fluctuation correlator
in a thermal gas of $\pi$-mesons to the lowest order in an
interaction due to a resonance exchange. A diagrammatic
technique is used. 
We discuss how this result can be applied
to event-by-event fluctuations in heavy-ion collisions,
in particular, to search for the critical point of QCD.
As a practical example, we determine 
the shape of the rapidity correlator.
\end{abstract}

\section{Introduction}

The goal of heavy-ion collision experiments is to shed
light on thermodynamic properties of strongly interacting matter and
 to determine the phase diagram of Quantum Chromodynamics
(QCD).  Recently, experimental \cite{na49,star,na49-new}
and theoretical 
\cite{C_V,Phi_pt,SRS1,SRS2,Berdnikov,moments,subevent,multiplicity,JK-res,charge,balance,HJ,Gazdzicki,SS,Hwa,Gavin,Mrow,Mrow-T} 
 study of
event-by-event fluctuations has attracted interest, in particular,
 because these
fluctuations carry information about QCD thermodynamics.
Recent experimental data from RHIC and CERN SPS 
\cite{star,na49-new} show non-trivial pattern of 
fluctuations which awaits theoretical understanding.

Thermodynamic description of the final stage of the
heavy-ion collision is motivated, to a large extent,
 by a remarkable phenomenological success
of statistical thermal model in describing
 the observed particle spectra \cite{PBM}.
As far as the fluctuations are concerned, the observed fluctuations are
Gaussian to a very good approximation, and are consistent with
being largely of thermodynamic origin (see
\cite{SRS2} for more quantitative analysis). In this paper
we shall not discuss the relationship between the thermodynamic
and observed fluctuations in heavy-ion collisions
 in detail (in particular, we shall not discuss possible importance of
non-equilibrium effects --- see, e.g., \cite{bubbling}).  At temperatures of
order 120 MeV characteristic of the final stage (freezeout)
of a high-energy heavy-ion collision, hot QCD is, 
to a good approximation, a gas of predominantly pions
interacting through exchange of resonances.
This motivates the main goal of this paper: to calculate equilibrium 
thermodynamic fluctuations given the
interaction of pions. 

The most interesting application of our results is the search 
for the critical point of QCD. As discussed in \cite{SRS1,SRS2}, such a
point is characterized by a divergent correlation length, i.e., a
massless excitation, in the channel with the quantum numbers of the
sigma meson (scalar isoscalar). The strategy of the search is to scan
the QCD phase diagram by varying experimental parameters, 
such as the energy of the collision,
looking for the signatures of the critical point \cite{SRS1,SRS2}. 
The fluctuations 
of pions 
 are sensitive to the presence of light sigma excitation
because of the direct coupling of pions and sigma: $\sigma\pi\pi$.
Thus, measuring the pion fluctuations one could determine if the 
freezeout is occuring near the critical point \cite{SRS2}.

Another important application of our results is the study of charge
fluctuations \cite{JK-res,charge,balance,HJ,Gazdzicki,SS}.

The purpose of the paper is more generic. We wish to derive a formula for the
pion fluctuations, specifically, for the two-pion correlator,
given the effective interactions of pions.
We shall discuss in detail the effect of the interaction induced
by the sigma exchange, both because of its application to
the search for the critical point and because this interaction
is the simplest. It is relatively straightforward to generalize
the analysis and the results to interactions induced by exchanges
of other resonances, such as $\rho$-meson.

The goal of this paper is to motivate and derive
formulas such as (\ref{G2corr}). As a simple example of an application
of these results we determine the rapidity correlator (\ref{Cyy}).
Other applications and a detailed
comparison of these results to experiment are deferred to further work.

\section{Two-particle correlator and fluctuations}

Most of the commonly used fluctuation measures, such as
Gaussian widths of event-by-event distributions of 
mean transverse momenta, $\langle(\Delta p_T)^2\rangle$, 
or correlation coefficients such as, e.g., $\langle\Delta p_T \Delta
N\rangle$ ($\Delta N$ is the multiplicity fluctuation), can be
related to the two-particle correlator, or the two-particle momentum
space density, as emphasized in \cite{SRS2,moments}.
For completeness, and to underline the importance of the
two-particle correlator, we shall repeat some typical relationships here.
An expert may wish to skip this section.

Let us denote by $\langle\ldots\rangle$ an average over events.
Let us assume that we binned the phase space into bins labeled
by the values of momenta $\bbox{p}$ (or $\bbox{k}$). We denote by 
$n_{\bbox{p}}$
the occupation number of a given bin in a given event. Then
$\Delta n_\p\equiv n_\p - \langle n_\p \rangle$ is the fluctuation,
in a given event, of this occupation number around the all-event mean.
The two-point fluctuation
correlator which will be discussed and calculated in this
paper is then given by
\begin{equation}\label{corr}
\langle \Delta n_{\bbox{p}}\, \Delta n_{\bbox{k}} \rangle.
\end{equation}

For example, the event-by-event fluctuation of the mean transverse momentum
is related to the correlator (\ref{corr}):
\begin{equation}\label{dpt2}
\langle(\Delta p_T)^2\rangle = {1\over\langle N\rangle^2}
\sum_{\bbox{p}}\sum_{\bbox{k}}
\langle \Delta n_{\bbox{p}}\, \Delta n_{\bbox{k}} \rangle
(p_T-\overline {p_T})(k_T-\overline {p_T}),
\end{equation}
where $\overline {p_T}$ is the inclusive mean transverse momentum.
Similarly, the correlator of $p_T$ and multiplicity fluctuations
(also known as linear correlation coefficient), can be expressed as
\begin{equation}\label{dptdN}
\langle \Delta p_T\Delta N\rangle = {1\over\langle N\rangle}
\sum_{\bbox{p}}\sum_{\bbox{k}}
\langle \Delta n_{\bbox{p}}\, \Delta n_{\bbox{k}} \rangle
(p_T-\overline {p_T}).
\end{equation}
The multiplicity fluctuation is related to (\ref{corr}) by an obvious
formula:
\begin{equation}\label{dN2}
\langle (\Delta N)^2\rangle =
\sum_{\bbox{p}}\sum_{\bbox{k}}
\langle \Delta n_{\bbox{p}}\, \Delta n_{\bbox{k}} \rangle.
\end{equation}

It is straightforward to generalize these relationships to
correlations of fluctuations of momenta or multiplicities of
particles of {\em different species}, e.g., $\pi^+$ and $\pi^-$.
The most interesting example is the charge fluctuation, $\Delta Q$,
which can be expressed via a correlator such as (\ref{mastercorr}):
\begin{equation}\label{Q2}
\langle (\Delta Q)^2\rangle =
\sum_{\alpha\bbox{p}}\sum_{\beta\bbox{k}}Q^\alpha Q^\beta\npnk,
\end{equation}
where $Q^\alpha$ is the charge of the particle labeled $\alpha$.

The correlator (\ref{corr}) also equals
$\langle n_{\bbox{p}}\,n_{\bbox{k}} \rangle
- \langle n_{\bbox{p}}\rangle\langle n_{\bbox{k}} \rangle$.
Since $\langle n_{\bbox{p}}\,n_{\bbox{k}} \rangle - \langle
n_{\bbox{p}}\rangle\delta_{\p\k}\equiv \rho_2(\p,\k)$ 
is the two-particle momentum space density, one can
see that the absence of correlations, i.e.,  
$\rho_2(\p,\k)=\rho(\p)\rho(\k)$ translates into
\begin{equation}\label{triv}
\langle \Delta n_{\bbox{p}}\, \Delta n_{\bbox{k}} \rangle = 
\langle
n_{\bbox{p}}\rangle\delta_{\p\k}.
\end{equation}
In the literature such trivial fluctuations
are termed ``statistical''. For example, with (\ref{triv}):
$\langle (\Delta N)^2\rangle = \langle N\rangle$, 
$\langle \Delta p_T\Delta N\rangle=0$, etc.  Corrections 
to (\ref{triv}) are commonly termed ``dynamical'' fluctuations. In this
paper we shall calculate corrections to (\ref{triv}) due to
interactions.

The relationships of the type (\ref{dpt2})-(\ref{dN2}) and (\ref{Q2}) 
demonstrate that the correlator
(\ref{corr}) is a very universal quantity.

\section{Calculation of the correlator}

\subsection{Correlator in an ideal gas}

Our goal is to calculate the universal fluctuation
correlator which we define as
\begin{equation}\label{mastercorr}
\langle \Delta n^\alpha_{\bbox{p}}\, \Delta n^\beta_{\bbox{k}} \rangle,
\end{equation}
where $\langle\ldots\rangle$ denotes the average over a thermodynamic
ensemble,
$\Delta n^\alpha_{\bbox{p}} =  n^\alpha_{\bbox{p}} - \langle
n^\alpha_{\bbox{p}}\rangle$ and $n^\alpha_{\bbox{p}}$ is the 
occupation number for a particle of type $\alpha$ in the 
momentum mode $\bbox{p}$. In view of experimental applications
we shall focus on charged pions, i.e, $\alpha=+,-$
for $\pi^+$ and $\pi^-$.

In the ideal Bose gas this correlator is given by a well-known
formula \cite{Landau}
\begin{equation}\label{freecorr}
\langle 
\Delta n^\alpha_{\bbox{p}}\, \Delta n^\beta_{\bbox{k}}
\rangle_{\rm free}
= \delta^{\alpha\beta}\delta_{\bbox{p}\bbox{k}}
f_{\bbox{p}}(1+f_{\bbox{p}}).
\end{equation}
where 
\begin{equation}
f_\p=\langle n^\alpha_\p\rangle = {1\over e^{\beta\omega_\p}-1}
\end{equation}
is equilibrium Bose-Einstein distribution function.
Eq. (\ref{freecorr})
simply means that occupation numbers in two different modes are
uncorrelated and in each mode the mean square variance is as in
the Poisson distribution, $\langle n_{\bbox{p}}\rangle$, 
times the Bose enhancement
factor, $\langle 1+ n_{\bbox{p}}\rangle$.
Without the Bose enhancement factor the correlator is
just the trivial correlator (\ref{triv}). The corrections due to
Bose enhancement, significant in the region of $\p$ and $\k$
very close to each other, are the subject of 
Hanbury-Brown-Twiss (HBT) interferometry studies and will not be
discussed here (see, e.g., \cite{Heinz}
for review).

It is useful to note that $f_\p(1+f_\p)$ in the r.h.s. of
(\ref{freecorr}) is the first derivative of $f(\beta\omega_\p)$:
\begin{equation}\label{f'}
f_\p' \equiv f'(\beta\omega_\p) = f_\p(1+f_\p).
\end{equation}


\subsection{Effect of interaction}

We wish to find the effect of the interaction on the
correlator (\ref{mastercorr}). 
Once the interaction is turned on we must carefully
rethink the meaning of $n_\p$. 
The spectrum is no longer a
direct superposition of the single-particle spectra, i.e., the energy
of a level is not just $\sum_\p n_\p \omega_\p$. In other words, the
scattering now makes occupation numbers of individual momentum modes
bad quantum numbers --- they are not conserved. However, if the
changes in the spectrum are controllably small, we may be able to trace the
shifted levels to their original position in the free gas and assign
their quantum numbers correspondingly. 
We shall see below how this qualitative picture works in the lowest
non-trivial order in the interaction. 

Such an ``adiabatic'' definition
of the occupation numbers would correspond to a situation in which
the interaction (i.e., scattering) in a thermalized system is turned off at a
point in time, the momenta are thus frozen, and the particles 
are allowed to stream freely into a detector surrounding the system.
Such a situation can be considered as a model of freezeout in heavy-ion
collisions. 

Such a picture relies on the weakness of the interaction and,
presumably, it cannot be pushed beyond the leading order calculation
systematically. However, we would like to apply it to the gas of
rather strongly interacting particles (such as pions).  We shall argue that
some of the contributions of interactions can be absorbed into
self-energy, or vertex type renormalizations. We then assume,
unfortunately without a proof, that the resulting effective
interactions, {\em to lowest order}, would be sufficient to determine
the correlator to a satisfactory (from experimental point of view)
precision, in the same way as a tree level effective Lagrangian
describes strong interaction amplitudes. Although our motivation
requires us to make such an assumption, in the remainder of this
section we shall simply concentrate on performing a formal lowest
order calculation, assuming that coupling constant can be made
controllably small.  We shall return to the discussion of applicability
of our results again.


\subsection{Interaction and free energy}

In summary, this is the strategy.
The energy of a level, descendent from a free level with occupation
numbers given by a set $\{n_\p\}$ is $E[\{n_\p\}]=\sum_\p n_\p
\omega_\p + \delta E[\{n_\p\}]$, where $\delta E$ 
is a correction which depends, in
principle, on all $n_\p$.  Once we know the spectrum $E[\{n_\p\}]$ we
can determine the joint probability distribution $\exp(-\beta E[\{n_\p\}])$
for the sets $\{n_\p\}$ and thus
calculate correlators such as (\ref{mastercorr}). 
\footnote{
One should note here
that such a calculation must break down if there is mixing between
levels which are degenerate without interaction (scattering or decay
of an unstable resonance is related to such a mixing). 
In this case, $\delta E$ becomes singular 
in the leading order in perturbation theory. This fact we shall
observe in our calculation.}

Rather than pushing forward along these lines, we shall, instead,
begin by a more familiar calculation of the free energy 
of the interacting gas using
perturbative expansion. We shall see that this calculation
can be mapped onto the scheme outlined above.

We consider a theory with two charged pions $\pi^+$ and $\pi^-$ of
mass $m$ and
a scalar field $\sigma$ of mass $\tilde m$, 
with the interaction controlled by a coupling
$G$ and given by the Lagrangian
\begin{equation}\label{LI}
{\cal L}_{\sigma\pi\pi} = 2G \sigma \pi^+ \pi^-. 
\end{equation}
We shall only concentrate on the effect of this simple interaction.
It will then be rather straightforward to include other fields and
their interactions. 

\begin{figure}
\centerline{
\epsfig{file=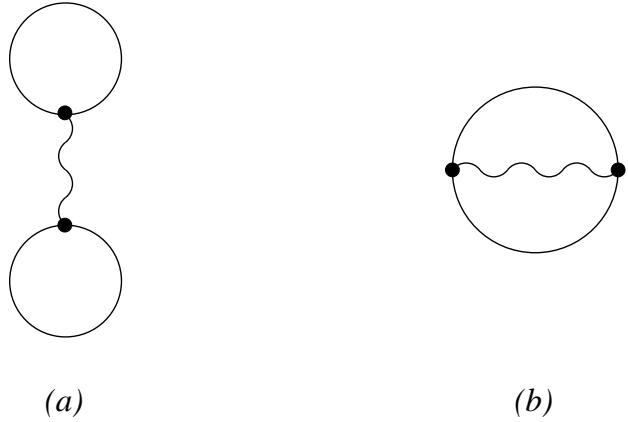, width=.5\textwidth}
}
\caption[]{Diagrams for the free energy at order $G^2$.}
\label{fig:diagrams}
\end{figure}

To order $G^2$, the correction to the free energy is given by the
two diagrams in Fig. \ref{fig:diagrams}. We write the
corresponding expressions using the mixed time-momentum representation
(see, e.g., \cite{Abrikosov,Weldon,Pisarski}) for the propagators:
\begin{equation}
\Delta(\tau,\p) = {1+f_\p\over2\omega_\p}\left(e^{-\omega_\p\tau} +
e^{-\omega_\p(\beta-\tau)}\right)\,.
\end{equation}
In this representation the integral  over the time $\tau$ separation of the
vertices in these diagrams is trivial. The 
result can be written as:
\begin{mathletters}
\label{F2}
\begin{equation}\label{F2a}
F^{(2a)} = - 2G^2 
\int_\p\int_\k
{1\over2\omega_\p}
{1\over2\omega_\k}
{1\over \tilde m^2}
(1 + 2f_\p)
(1 + 2f_\k)\,,
\end{equation}
\begin{equation}\label{F2b}
F^{(2b)} = - 4G^2 
\int_\p\int_\k
{1\over2\omega_\p}
{1\over2\omega_\k}
{1\over2\omega_\q}
\sum_{\{s_i=\pm1\}}
{1\over\sum_i\omega_i s_i}\left[
\prod_i f_i^{1-s_i\over2}(1+f_i)^{1+s_i\over2}
 \right],
\end{equation}
\end{mathletters}
where
$\int_\p = \int d^3\p/(2\pi)^3$, 
$\q=\p+\k$. The first expression, $F^{(2a)}$, is straightforward, while
for  $F^{(2b)}$ we introduced shorthand notations:
 $i=1,2,3$, $\omega_1=\omega_\p$, $\omega_2=\omega_\k$, 
$\omega_3=\tilde\omega_\q$, and analogous notations for $f_i$'s. 
To elucidate the expression (\ref{F2b}), we write one term from the 8
terms in the sum over sets $\{s_i\}$ explicitly. The term
\begin{equation}\label{term}
- 4G^2 
\int_\p\int_\k
{1\over2\omega_\p}
{1\over2\omega_\k}
{1\over2\omega_\q}
{1\over\omega_\p+\omega_\k+\tilde\omega_\q}
(1+f_\p)(1+f_\k)(1+\tilde f_\q) 
\end{equation}
corresponds to $\{s_i\}=\{1,1,1\}$.
The meaning of this term is simple. It is the ensemble average of the
second order perturbative correction to the energy,
since to  leading order 
\begin{equation}
\delta F = \langle \delta E\rangle \equiv 
\sum_A \delta E_A e^{-\beta E_A}/Z. 
\end{equation}
The sum $\omega_\p+\omega_\k+\tilde\omega_\q$ in (\ref{term})
is the energy denominator.
The term  exhibited in (\ref{term}) comes from the virtual intermediate
state $A'=\{n_\p^++1,n_\k^-+1,\tilde n_\q+1\}$, 
which has three extra particles $\pi^+$, $\pi^-$ and $\sigma$,
relative to a given state $A=\{n_\p^+,n_\k^-,\tilde n_\q\}$. 
All terms in (\ref{F2b}) correspond to virtual transitions
generated by $2G\sigma\pi^+\pi^-$. There are 8 variants:
a state with either one more {\em or} one fewer $\sigma$
($s_3=+1$ or $-1$), {\em and}
either one more $\pi^+$ {\em or} one fewer $\pi^-$
($s_1=\pm1$), {\em and}
either one more $\pi^-$ {\em or} one fewer $\pi^+$
($s_2=\pm1$). The
first, simpler, diagram, $F^{(2a)}$, describes the contribution of
the virtual states with one more/fewer $\sigma$, and unchanged 
numbers of $\pi^+$ and $\pi^-$.


We see that the energy of a given state characterized
by a set of occupation numbers $A=\{n_\p^+,n_\k^-,\tilde n_\q\}$, 
receives a $G^2$ correction $\delta E[\{n^+_\p, n^-_\k, \tilde n_\q\}]$,
which is a function(al) of the occupation numbers $A$ in the
unperturbed state. 
Thus, at least up to order $G^2$, we can view the energy $E_A=E_A^{(0)}+
\delta E_A$ of a 
given state as a function(al) of the occupation numbers in this state. 
This allows us to evaluate thermodynamic averages such 
as (\ref{mastercorr}).

\subsection{The correlator}

At zeroth order (i.e., 
in the free theory), since  $E_A^{(0)}$ is linear in $n_\p$, 
$\exp(-\beta E_A^{(0)})$ factorizes into Gibbs distributions for individual
momentum modes, and the correlator is given by (\ref{freecorr}).
However, 
the correction $\exp(-\beta \delta E[\{n^+_\p, n^-_\k, \tilde n_\q\}])$ 
does not factorize, introducing correlations between occupation
numbers of different modes.

\begin{figure}
\centerline{
\epsfig{file=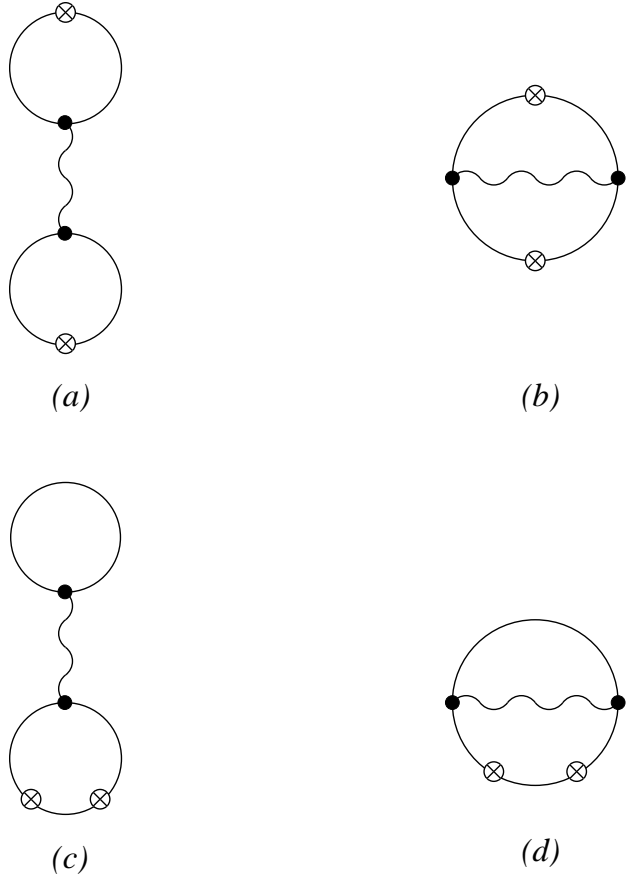,width=.5\textwidth}
}
\caption[]{Diagrams for the correlator}
\label{fig:xxdiagrams}
\end{figure}

One can derive simple rules which allow to read off the
correlator directly from the expression for the free energy.
To facilitate this, one can introduce ``chemical potentials''
for relevant occupation numbers, shifting the energy
$E[\{n^+_\p,n^-_\k,\tilde n_\q\}]$ by
$-\sum_\alpha\sum_\p \mu^\alpha_\p n^\alpha_\p$.
Differentiating resulting $\mu$-dependent free energy 
with respect to $\mu^\alpha_\p$ and $\mu^\beta_\k$ we obtain correlators
such as (\ref{mastercorr}). In the zeroth order this, of course, reproduces
(\ref{freecorr}). In the order $G^2$ the result can be represented by
diagrams in Fig. \ref{fig:xxdiagrams}. These diagrams are obtained from those
in Fig. \ref{fig:diagrams}. A cross represents differentiation with
respect to  $\mu^\alpha_\p$ and amounts to replacing corresponding
$f^\alpha_\p$
or $(1+f^\alpha_\p)$ factors with $f^\alpha_\p(1+f^\alpha_\p)$,
which is the derivative of $f^\alpha_\p$ with respect to
$\beta\omega_\p$ (see (\ref{f'})).
\footnote{Although, in equilibrium, $f^+_\p$ and $f^-_\p$ are the
same, we are using different
notations for them since they differ when $\mu^+_\p$ and $\mu^-_\p$ 
are introduced.}

There is a class of diagrams in Fig.~\ref{fig:xxdiagrams} which affect the
correlator (\ref{mastercorr}) in a rather trivial way. These
are the diagrams (c) and (d), 
which have both crosses on the same propagator line.
Such diagrams give contribution proportional to 
$\delta^{\alpha\beta}\delta_{\p\k}$. These diagrams represent
$O(G^2)$ self-energy type corrections $\delta \omega_\p$ 
to $\omega_\p$ multiplied by a factor
$f_\p{''}$. Since at zeroth order the correlator is 
$\delta^{\alpha\beta}\delta_{\p\k}f{'}(\beta\omega_\p)$ (\ref{freecorr}),
one can see that such diagrams are accounted for when we replace
$\omega_\p$ by $\omega_\p+\delta \omega_\p$ in $f{'}(\beta\omega_\p)$.

We shall not study the effects of such self-energy contributions
on the correlator or on the single-particle distributions $\langle
n_\p\rangle$.%
\footnote{It is useful to note that $O(G^2)$ corrections
to $\langle n_\p\rangle$ can be also calculated in the same way.
These are given by the same diagrams but with one cross only,
i.e., proportional to $f{'}_\p$.
One sees that these corrections are accounted for by
replacing $\omega_\p$ with its $O(G^2)$-corrected value in the
function $f(\beta\omega_\p)$.}
 Instead, we assume that these corrections have been
already accounted for and we are using corrected $\omega_\p$.
We can view this procedure in the sense of the Wilsonian
effective theory approach, and consider the Lagrangian with
interaction (\ref{LI}) as the effective Lagrangian. That means
no loops (such as self-energy) should be computed with it.%
From a practical point of view, one can imagine that $\langle
n_\p\rangle$ is measured rather than calculated, and the
results are expressed in terms of such measured distribution,
instead of~$f_\p$.

\begin{figure}
\centerline{
\epsfig{file=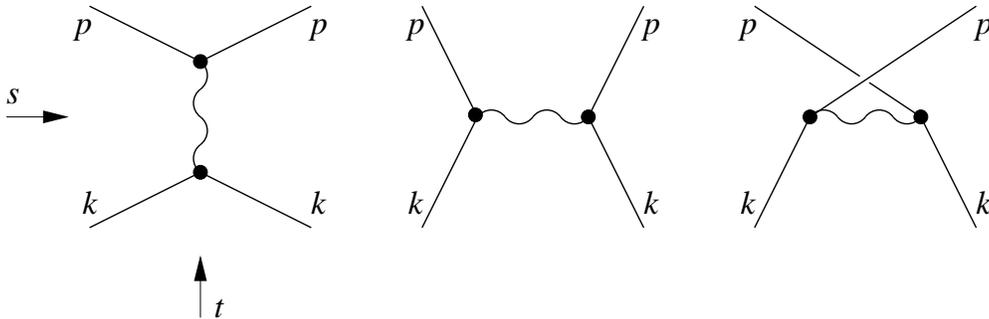,width=.8\textwidth}
}
\caption[]{A representation of 3 terms in the correlator
(\ref{G2corr}). The resonance is exchanged in $t$, $s$ or $u$ channel.}
\label{fig:tsu}
\end{figure}

With this understanding we can now focus on the most interesting
contributions to the correlator, given by the diagrams in 
Fig.~\ref{fig:xxdiagrams}(a,b). Differentiating (\ref{F2})
with respect to $\mu^\alpha_\p$ and $\mu^\beta_\k$, and
discarding self-energy terms proportional to $\delta_{\p\k}$,
we find the order $G^2$ correction to the correlator:
\begin{equation}\label{G2corr}
\npnk_{G^2} = 
\beta G^2 {f_\p' f_\k'\over\omega_\p\omega_\k}\left[
{\delta^{\alpha\beta}+\gamma^{\alpha\beta}\over\tilde m^2}
+
{\gamma^{\alpha\beta}\over\tilde\omega_\q^2-(\omega_\p+\omega_\k)^2}
+
{\delta^{\alpha\beta}\over\tilde\omega_\q^2-(\omega_\p-\omega_\k)^2}
\right]\,,
\end{equation}
where $f_\p'=f_\p(1+f_\p)$, $\gamma^{+-}=\gamma^{-+}=1$,
$\gamma^{++}=\gamma^{--}=0$.
 The first term in the square brackets is
from the diagram in Fig.~\ref{fig:xxdiagrams}(a), while the second and the
third are from Fig.~\ref{fig:xxdiagrams}(b).
A simple way to visualize the $\alpha\beta$ and $\p\k$ dependence in
these three terms
 is to consider a process of forward scattering
$\pi^\alpha_\p\pi^\beta_\k\to\pi^\alpha_\p\pi^\beta_\k$. The three
terms arise from contributions with the sigma propagator in $t$, $s$
and $u$ channels respectively, as illustrated in Fig.~\ref{fig:tsu}.

\section{Summary and discussion}

In this paper we considered pion gas in thermal equilibrium
as an idealization of the final (freezeout) stage of a
heavy-ion collision. We considered the effect of pion interaction
on the fluctuation correlator (\ref{corr}) or (\ref{mastercorr})
 --- an important universal
quantity characterizing pion fluctuations. 
Eq. (\ref{G2corr}) gives the lowest order correction
to the correlator (\ref{mastercorr}) due to the interaction
mediated by the exchange of $\sigma$-resonance and given
by the Lagrangian (\ref{LI}). Our main motivation to consider
such an interaction is the search for signatures of
the critical point on the QCD phase diagram, the point where the 
sigma mass vanishes. One can see that the first term in
(\ref{G2corr}), the one due to the sigma exchange in $t$-channel,
dominates in the limit $\tilde m\to0$. This term has been already
calculated in \cite{SRS2} using a different method. The advantage of the
diagrammatic approach of this paper is that it allows one to
obtain other terms (due to $s$- and $u$-channel exchange), 
subleading in the regime $\tilde m\to0$.

Strictly speaking, we have calculated the correction
to the correlator only in the leading order of perturbative
expansion. Thus, our results are only rigorous in the limit
of small coupling constant. Our motivation, however, has
been the calculation of a two-particle correlator in 
hot QCD,
which, at temperatures of around 120 MeV, characterizing the
freezeout stage of the heavy-ion collisions, is primarily 
a gas of pions, whose interactions are not controllably small.
We observed that,
at the leading order, some of the contributions could
be naturally absorbed into the self-energy renormalizations. 
We thus conjectured
that if one uses an {\em effective}  (rather than
bare) interaction in (\ref{LI}) and effective self-energies, 
our leading order calculation could be 
sufficient in practice.

Below we make several comments and
outline some questions and problems, related to the
result (\ref{G2corr}) and its comparison to experimental data.
 More detailed study of these questions is left
to further work.

\subsection{Other resonances}

The interaction due to the $\sigma$ meson exchange is simpler
than interaction due to other resonances, such as $\rho$,
but generalization is not difficult. For example,
given the effective $\rho^0\pi^+\pi^-$ interaction in the form
\footnote{We should remember that the tree-level interaction
which determines the pion correlator is the {\em effective}
interaction, which should includes all thermal self-energy and
vertex corrections. Therefore, the magnitude of the coupling
is not necessarily the same as in vacuum (although we do not expect it
to be significantly different in the case of $\rho$-meson), 
and the Lorentz invariance, which we
assume for simplicity, is not exact.}
\begin{equation}\label{LIrho}
{\cal L}_{\rho\pi\pi} = ig_{\rho\pi\pi} (\rho^0)^\mu \pi^+\partial_\mu\pi^- 
+ {\rm h.c.}, 
\end{equation}
one obtains the contribution of the $\rho$-exchange in the $t$-channel
in the form
\begin{equation}\label{rhocorr}
\npnk_{g^2,t-{\rm channel}} = 
\beta g_{\rho\pi\pi}^2 (p\cdot k) {f_\p' f_\k'\over\omega_\p\omega_\k}
{\delta^{\alpha\beta}-\gamma^{\alpha\beta}\over\tilde m_\rho^2}.
\end{equation}
Two notable differences from the similar $t$-exchange contribution
of $\sigma$
in (\ref{G2corr}) are worth a notice. Both are due to $\rho$
being a {\em vector} particle. First, there is an
{\em angular} correlation between the pions due to the $(p\cdot k)$ factor.
Second, the sign of the correlation now depends on the relative sign of
the two pions, due to a minus in 
$\delta^{\alpha\beta}-\gamma^{\alpha\beta}$.

Similar formulas can be derived for other resonance-mediated
interactions of pions. It should be clear that the contribution
of the $t$-exchange of $\sigma$-meson is dominant when
$m_\sigma\to0$,
i.e., near the critical point. It can be further distinguished
from other resonance contributions, such as $\rho$, by its
charge-independent character
($\delta^{\alpha\beta}+\gamma^{\alpha\beta}=1$
for all $\alpha\beta$). Most importantly, this contribution
will show a non-monotonic behavior if the
critical point is approached and then passed during the scan of the
QCD phase diagram.


\subsection{Resonance exchange vs decay}
It is important to realize the difference between the
result (\ref{G2corr}) and a more common approach to the
correlations due to resonance {\em decays} after freezeout \cite{SRS2,JK-res}.
It is easiest to see this difference in the charge dependence of the
correlator (\ref{G2corr}). The term which has the same charge
dependence as the contribution of the resonance decays is the
one due to $s$-channel exchange: $\sigma$ , or $\rho^0$,
decays into $\pi^+\pi^-$, therefore the correlator
must be proportional to $\gamma^{\alpha\beta}$. The contribution of $t$
and $u$ channel exchanges is entirely different.
In order to compare to experiment, however, one has to
take into account the fact that the {\em detected} pions consist to
a large extent (more than a half) of the products of resonance decays.
The contribution of such {\em resonance} 
pions to the correlator should be calculated
separately, using the kinematics of the resonance decays. Such 
a calculation 
has been done numerically, e.g., in \cite{SRS2} for $p_T$ fluctuations,
where the effect turned out to be small, except for the
contribution of the decay of light $\sigma$ meson near the critical point,
and in \cite{JK-res} for charge fluctuations, where the effect is of
order 20-30\%.
The emphasis of this paper is on the correlation induced by 
virtual resonance exchange among the {\em direct} pions.

\subsection{$s$-channel pole and resonance width}

Generalization to heavier resonances, such as $\rho$, presents another
problem.
If we consider $\p$ and $\k$ such that their invariant mass
 is equal to $\tilde m$, the second term in (\ref{G2corr}) becomes
infinite (the $s$-channel resonance is on mass shell). 
The perturbative expansion breaks down.%
\footnote{This is due to the degeneracy or level-crossing
as mentioned in an earlier footnote.
Unperturbed energy levels, different by
replacing $\rho$ with a $\pi^+\pi^-$ pair, 
 are degenerate in this case.}
However,
we know that there is no true pole singularity in this case, since the
resonance has a finite lifetime. Thus one would expect that
careful treatment of higher order terms should remove this
pole divergence and replace it with a smooth Breit-Wigner-type curve.
Perhaps, it will also allow to treat resonance exchange and decay
within the same formalism (cf. previous subsection).

The $s$-channel divergence
does not arise in the application to light $\sigma$
particle near the critical point, since in this case $m_\sigma<2 m_\pi$.
\footnote{However, higher order {\em thermal} 
corrections still induce imaginary part of the $\sigma$ propagator.}

\subsection{Longitudinal expansion and rapidity correlator}

In this paper we considered an isotropic pion gas (a fireball)
at rest. Before one
could compare our results to fluctuations observed in heavy-ion
collisions it is necessary to take into account the effect of
expansion. In this paper we shall only consider longitudinal
expansion. It can be taken into account
using Bjorken's boost-invariant hydrodynamic model. In this approximation,
one considers a superposition of fireballs placed continuously
over large (ideally infinite) rapidity interval. 
In order to obtain the correlator in this case one needs to
take the correlator for a gas at rest $\npnk_{\rm rest}$, 
boost momenta $\p$ and $\k$ with rapidity $y$: 
${\langle \Delta n^\alpha_{\bbox{p}\oplus y}\, \Delta
n^\beta_{\bbox{k}\oplus y}
\rangle}_{\rm rest}$ and integrate over $y$:
\begin{equation}\label{long}
\npnk_{\rm long.exp.}\equiv
\int_{-\infty}^{+\infty}dy\,
\omega_{\p\oplus y}\,\omega_{\k\oplus y}\,
{\langle \Delta n^\alpha_{\bbox{p}\oplus y}\, \Delta n^\beta_{\bbox{k}\oplus y}
\rangle}_{\rm rest}.
\end{equation}
The factors $\omega_\p$, $\omega_\k$ appear because
of the non-uniform density of modes in rapidity space, i.e., 
$d^3\p=\omega_\p\,d^2\p_T\,dy_\p$ ($y_\p={\rm atanh}(p_z/\omega_\p)$,
$\p_T=(p_x,p_y)$).

As a practical example, one can use 
formula (\ref{long}) to determine the form of rapidity
correlations. For this purpose one needs to fix rapidities
$y_\p$ and $y_\k$,
and integrate over transverse components $\p_T$ and $\k_T$:
\begin{equation}\label{Cyy}
C^{\alpha\beta}(y_\p,y_\k) = 
{dN/dy\over\int_\p f_\p}
\int_{\p_T} \int_{\k_T} \npnk_{\rm long.exp.},
\end{equation}
where $\int_{\p_T}\equiv\int d^2\p_T/(2\pi)^3$.
The normalization factor, $(dN/dy)/(\int_\p f_\p)$,
is such that $C(y_\p,y_\k)=\rho_2(y_\p,y_\k)-\rho(y_\p)\rho(y_\k)$,
where $\rho=dN/dy$ and $\rho_2$ is the density of pairs per unit
rapidity squared. 

\begin{figure}
\centerline{
\epsfig{file=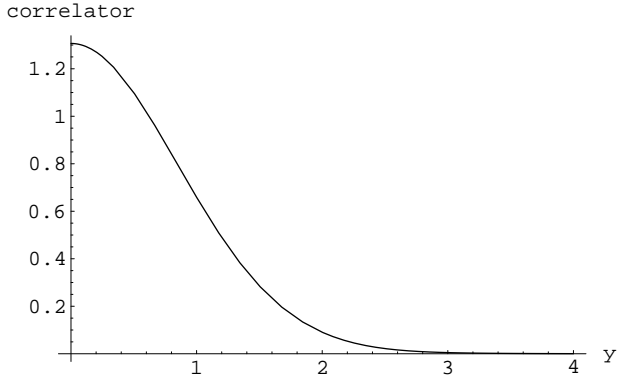,width=.5\textwidth}
}
\caption[]{A typical contribution to rapidity correlator. 
$y\equiv |y_\p-y_\k|$. In this example,
 $T=120$ MeV, $G/\tilde m=4.5$
(see \cite{SRS2,Berdnikov} for motivation). 
Plotted is the correlator $C$ in (\ref{Cyy}) divided by $dN/dy$.}
\label{fig:c}
\end{figure}

Using expression (\ref{long}) in
(\ref{Cyy}) one finds, of course, that correlator $C(y_\p,y_\k)$ is
a function of $y_\p-y_\k$ only. The dependence on $y_\p-y_\k$
can be found substituting (\ref{G2corr}) for $\npnk_{\rm rest}$
in (\ref{long}).
The detailed form of the correlator depends on the parameters
$G$, $\tilde m$, $\beta=1/T$. A rather robust feature of
$C(y_\p-y_\k)$, however, is that it vanishes exponentially with increasing
$|y_\p-y_\k|$ on the scale of 1-2 units. This is because
of the fast exponential falloff of the occupation number
factors $f_\p f_\k$ in (\ref{G2corr})
when either $\p$ or $\k$ becomes large compared
to $T$. As an example, the contribution to the rapidity 
correlator from the $t$-channel exchange term only, i.e., 
first term in (\ref{G2corr}), is shown in Fig.~\ref{fig:c}.

Another interesting consequence of (\ref{G2corr}) is a specific
dependence on multiplicity $dN/dy$. For example,
the centrality dependence of the freezeout parameters is known
to be rather weak. Neglecting it, we could predict, using
(\ref{G2corr}), that the centrality dependence comes only
from the factor $dN/dy$, and is, therefore, linear.

\section*{Acknowledgements}

The author thanks E.\ Shuryak, D.\ Son, M.\ Tsypin and A.\ Zhitnitsky for
discussions and comments and M.\ Gazdzicki, T.\ Trainor and S.\ Voloshin
for explaning to him many relevant  experimental issues. 
He also thanks RIKEN, Brookhaven National
Laboratory, and U.S.\ Department of Energy 
[DE-AC02-98CH10886] 
for providing the facilities essential for the completion of this
work. This work is supported, in part, by DOE OJI grant.


\begin{thebibliography}{99}




\bibitem{na49}
H.~Appelshauser {\it et al.}  [NA49 Collaboration],
Phys.\ Lett.\ B {\bf 459}, 679 (1999);
J.~G.~Reid  [NA49 Collaboration],
Nucl.\ Phys.\ A {\bf 661}, 407 (1999);
S.~V.~Afanasev {\it et al.}  [NA49 Collaboration],
Phys.\ Rev.\ Lett.\  {\bf 86}, 1965 (2001);
J.~Bachler {\it et al.}  [NA49 Collaboration],
Nucl.\ Phys.\ Proc.\ Suppl.\  {\bf 92}, 7 (2001).


\bibitem{star}
J. Reid [STAR Collaboration], in proceedings of 
15th International Conference on Ultrarelativistic 
Nucleus-Nucleus Collisions (QM2001),
Stony Brook, New York, 15-20 Jan 2001; 
S.~A.~Voloshin [STAR Collaboration],
arXiv:nucl-ex/0109006.

\bibitem{na49-new}
V. Friese {\it et al.} [NA49 Collab], in proceedings of
International Symposium on Multiparticle Dynamics,
Datong, China, Sept. 1-7,2001, to be published by 
World Scientific.




\bibitem{C_V}
L.~Stodolsky,
Phys.\ Rev.\ Lett.\  {\bf 75}, 1044 (1995);
E.~V.~Shuryak,
Phys.\ Lett.\ B {\bf 423}, 9 (1998).



\bibitem{Phi_pt}
M.~Gazdzicki and S.~Mrowczynski,
Z.\ Phys.\ C {\bf 54}, 127 (1992);
M.~Gazdzicki, A.~Leonidov and G.~Roland,
Eur.\ Phys.\ J.\ C {\bf 6}, 365 (1999).



\bibitem{SRS1}
M.~Stephanov, K.~Rajagopal and E.~V.~Shuryak,
Phys.\ Rev.\ Lett.\  {\bf 81}, 4816 (1998).

\bibitem{SRS2}
M.~Stephanov, K.~Rajagopal and E.~V.~Shuryak,
Phys.\ Rev.\ D {\bf 60}, 114028 (1999).

\bibitem{Berdnikov}
B.~Berdnikov and K.~Rajagopal,
Phys.\ Rev.\ D {\bf 61}, 105017 (2000).






\bibitem{moments}
A.~Bialas and V.~Koch,
Phys.\ Lett.\ B {\bf 456}, 1 (1999);
M.~Belkacem, Z.~Aouissat, M.~Bleicher, H.~Stocker and W.~Greiner,
arXiv:nucl-th/9903017.

\bibitem{subevent}
S.~A.~Voloshin, V.~Koch and H.~G.~Ritter,
arXiv:nucl-th/9903060.

\bibitem{multiplicity}
G.~Baym and H.~Heiselberg,
Phys.\ Lett.\ B {\bf 469}, 7 (1999).
G.~V.~Danilov and E.~V.~Shuryak,
arXiv:nucl-th/9908027.

\bibitem{JK-res}
S.~Jeon and V.~Koch,
Phys.\ Rev.\ Lett.\  {\bf 83}, 5435 (1999).



\bibitem{charge}
M.~Asakawa, U.~W.~Heinz and B.~Muller,
Phys.\ Rev.\ Lett.\  {\bf 85}, 2072 (2000);
S.~Jeon and V.~Koch,
Phys.\ Rev.\ Lett.\  {\bf 85}, 2076 (2000);
M.~Bleicher, S.~Jeon and V.~Koch,
Phys.\ Rev.\ C {\bf 62}, 061902 (2000).

\bibitem{balance}
S.~A.~Bass, P.~Danielewicz and S.~Pratt,
Phys.\ Rev.\ Lett.\  {\bf 85}, 2689 (2000).

\bibitem{HJ}
H.~Heiselberg and A.~D.~Jackson,
Phys.\ Rev.\ C {\bf 63}, 064904 (2001).

\bibitem{Gazdzicki}
M.~Gazdzicki and S.~Mrowczynski,
arXiv:nucl-th/0012094.

\bibitem{SS}
E.~V.~Shuryak and M.~A.~Stephanov,
Phys.\ Rev.\ C {\bf 63}, 064903 (2001).

\bibitem{Hwa}
R.~C.~Hwa and C.~B.~Yang,
arXiv:hep-ph/0104216.


\bibitem{Gavin}
D.~Bower and S.~Gavin,
Phys.\ Rev.\ C {\bf 64}, 051902 (2001).



\bibitem{Mrow}
S.~Mrowczynski,
Phys.\ Lett.\ B {\bf 465}, 8 (1999);
Acta Phys.\ Polon.\ B {\bf 31}, 2065 (2000).

\bibitem{Mrow-T}
R.~Korus, S.~Mrowczynski, M.~Rybczynski and Z.~Wlodarczyk,
arXiv:nucl-th/0106041.






\bibitem{PBM}
P.~Braun-Munzinger, J.~Stachel, J.~P.~Wessels and N.~Xu,
Phys.\ Lett.\ B {\bf 344}, 43 (1995);
Phys.\ Lett.\ B {\bf 365}, 1 (1996);
P.~Braun-Munzinger, I.~Heppe and J.~Stachel,
Phys.\ Lett.\ B {\bf 465}, 15 (1999).




\bibitem{bubbling}
K.~Rajagopal,
Nucl.\ Phys.\ A {\bf 680}, 211 (2000).


\bibitem{Landau} L. D. Landau and E. M. Lifshitz, {\em Statistical Physics},
Part 1 (Pergamon, 1980).

\bibitem{Heinz}
U.~W.~Heinz and B.~V.~Jacak,
Ann.\ Rev.\ Nucl.\ Part.\ Sci.\  {\bf 49}, 529 (1999).


\bibitem{Abrikosov}
A.A. Abrikosov, L.P.Gorkov and I.E. Dzyaloshinski,
{\em Methods of quantum field theory in statistical physics},
(Dover, New York, 1975).


\bibitem{Weldon}
H.~A.~Weldon,
Phys.\ Rev.\ D {\bf 28}, 2007 (1983).

\bibitem{Pisarski}
R.~D.~Pisarski,
Nucl.\ Phys.\ B {\bf 309}, 476 (1988).


\bibitem{BJ}
J.~D.~Bjorken,
Phys.\ Rev.\ D {\bf 27}, 140 (1983).

\end{thebibliography}
\end{document}